\documentclass[conference]{IEEEtran}
\IEEEoverridecommandlockouts
\usepackage[dvipdfmx]{graphicx}
\usepackage{cite}
\usepackage{textcomp}
\usepackage{amsmath,amssymb,amsfonts}
\usepackage{textcomp}
\usepackage{xcolor}
\usepackage{bm}
\usepackage{amsbsy}
\usepackage{color}
\usepackage[utf8]{inputenc} 
\usepackage[T1]{fontenc}  
\usepackage{xcolor}
\usepackage{hyperref}       
\hypersetup{
    colorlinks,%
    citecolor=black,%
    filecolor=black,%
    linkcolor=black,%
    urlcolor=black
}
\usepackage{url}           
\usepackage{booktabs}      
\usepackage{nicefrac}      
\usepackage{microtype}      
 \usepackage{algorithm}
 \usepackage[noend]{algpseudocode}
\usepackage{bbm}
\usepackage{mathtools}

\DeclareMathOperator*{\argmax}{argmax}

\bgroup
\usepackage[acronym,toc,shortcuts]{glossaries}
\makeglossaries
\newacronym{DDQN}{DDQN}{deep double Q-network}
\newacronym{DRL}{DRL}{deep reinforcement learning} 
\newacronym{DRQ}{DRQ}{dual reinforcement Q-routing}
\newacronym{DQN}{DQN}{deep Q-networks}
\newacronym{DNN}{DNN}{deep neural network}
\newacronym{IS}{IS}{importance-sampling}
\newacronym{IP}{IP}{internet protocol}
\newacronym{SARSA}{SARSA}{state–action–reward–state–action}
\newacronym{SDN}{SDN}{software-defined network}
\newacronym{TD}{TD}{temporal-difference}
\newacronym{TE}{TE}{traffic engineering}
\newacronym{ReLU}{ReLU}{rectified linear unit}
\newacronym{RL}{RL}{reinforcement learning}
\newacronym{PCE}{PCE}{path computation element}
\newacronym{PER}{PER}{prioritized experience replay} 
\newacronym{QoS}{QoS}{quality of service}

\begin{document}
\title{Hierarchical Deep Double Q-Routing}

 \author{
	\IEEEauthorblockN{Ramy E. Ali \IEEEauthorrefmark{1}\IEEEauthorrefmark{2}, Bilgehan Erman
		\IEEEauthorrefmark{2}, Ejder Ba\c{s}tu\u{g}
		\IEEEauthorrefmark{3} and Bruce Cilli
		\IEEEauthorrefmark{2}
	}
	\IEEEauthorblockA{\IEEEauthorrefmark{1}%
		Electrical Engineering Department, Penn State University, University Park, PA, USA,
		E-mail: \href{mailto:ramy.ali@psu.edu}{ramy.ali@psu.edu} }
	\IEEEauthorblockA{\IEEEauthorrefmark{2}%
		Nokia Bell Labs, Murray Hill, NJ, USA, 
		E-mails: \href{mailto:bilgehan.erman@nokia-bell-labs.com}{bilgehan.erman@nokia-bell-labs.com},  \href{mailto:bruce.cilli@nokia-bell-labs.com}{bruce.cilli@nokia-bell-labs.com}} 
	\IEEEauthorblockA{\IEEEauthorrefmark{3}%
	Nokia Bell Labs, Paris-Saclay, France, E-mail: \href{mailto:ejder.bastug@nokia-bell-labs.com}{ejder.bastug@nokia-bell-labs.com}
	}
}
\maketitle
%%%%%%%%%%%%%%%%%%%%%%%%%%%%%%%%%%%%%%%%%%%%%%%%%%%%%%%%%%%%%%%%%%%%%%%%
%%%%%%%%%%%%%%%%%%%%%%%%%%%%%%%%%%%%%%%%%%%%%%%%%%%%	ABSTRACT
\begin{abstract}
This paper explores a deep reinforcement learning approach applied to the packet routing problem with high-dimensional constraints instigated by dynamic and autonomous communication networks. Our approach is motivated by the fact that centralized path calculation approaches are often not scalable, whereas the distributed approaches with locally acting nodes are not fully aware of the end-to-end performance. We instead hierarchically distribute the path calculation over designated nodes in the network while taking into account the end-to-end performance. Specifically, we develop a hierarchical cluster-oriented adaptive per-flow path calculation mechanism by leveraging the Deep Double Q-network (DDQN) algorithm, where the end-to-end paths are calculated by the source nodes with the assistance of cluster (group) leaders at different hierarchical levels. In our approach, a deferred composite reward is designed to capture the end-to-end performance through a feedback signal from the source node to the group leaders and captures the local network performance through the local resource assessments by the group leaders. Our approach is expected to scale in large networks, adapt to the dynamic demand, utilize the network resources efficiently and can be applied to segment routing. 
\end{abstract}
\begin{IEEEkeywords}
hierarchical routing, dynamic routing, segment routing, path calculation, deep double Q-network
\end{IEEEkeywords}
%
%%%%%%%%%%%%%%%%%%%%%%%%%%%%%%%%%%%%%%%%%%%%%%%%%%%%%%%%%%%%%%%%%%%%%%%%
%%%%%%%%%%%%%%%%%%%%%%%%%%%%%%%%%%%%%%%%%%%%%%%%%%%%	INTRODUCTION
\section{Introduction}
Routing is one of the most challenging problems in \ac{IP} packet networking, with the general form of its optimization being interpreted as the NP-complete multi-commodity flow problem \cite{di1998efficient}. The key factors contributing to the routing complexity are the large number of concurrent demands, each with specific desired \ac{QoS} constraints, and limited shared resources in the form of number of links and their limited capacities. In large-scale networks with high-dimensional end-to-end \ac{QoS} constraints, the routing algorithm requires incorporation of additional attributes such as delay, throughput, packet loss, network topology and other \ac{TE} factors. On top of that, the recent transformation of the \ac{IP} networking towards a virtualized architecture and an autonomous control plane further increased the routing challenge by incorporating additional factors including the live status from the lower layers instances and the higher level network slice instance  such as the estimated duration of the flow. This network paradigm introduces more dynamism to the network operation as the traffic profiles in such scenarios are subject to change in much smaller time-scales, and given such a dynamism in hand, a significantly short reaction time is required which results in significant cost increase in the  path computation making it harder for a central control to compute all real-time routes in a large-scale network.

Today, source routing methods such as segment routing \cite{filsfils2015segment} in conjunction with \glspl{PCE} \cite{ash2006path} provide much-needed capabilities to resolve many of these routing challenges, by  allowing intermediate nodes to perform routing decisions thus offloading centralized path computation entities of the network. The reliance on central control nodes still stands as an open issue as the network scales, therefore, motivating us in to study and find a balance between centralized-vs-decentralized operation. In this paper, we propose a hierarchical routing scheme leveraging the recent advances in \ac{DRL}, to manage the complexities of routing problem. In order to distribute the path computation load in large-scale networks while taking into account the end-to-end performance, we develop a hierarchical \ac{QoS}-aware per-flow path computation algorithm. In our approach, the nodes in the network are grouped (clustered), based on specific criteria such as the latency between the nodes, with each group having a designated leader assigned either autonomously or predefined by the network operator. The end-to-end route from a source node to a destination is calculated by the source node with the assistance of group leaders at different hierarchical levels. The group leaders select links based on the local information available such as link utilizations, topology information and delays while taking into account the global view of the network through the feedback of the source nodes. This feedback signal of the source node represents the source's satisfaction with the calculated path. 

In our approach, while the stateful computations are done at the source nodes, the assisting group leaders behave as stateless functions for per-route computation threads. The dynamically computed routes are directly applicable to segment routing for establishment of the packet flows. The group leaders in this regard provide assistance to route calculation based on their local network conditions, while the source nodes maintains responsibility of assembling the end-to-end route.

The remainder of this paper is organized as follows. Section \ref{Background and Related Work} provides a brief overview about \ac{RL} and the related work applying \ac{RL} in the routing problem. Section \ref{Problem Formulation} includes our formulation of the hierarchical routing problem and our \ac{DRL}-enabled routing algorithm. We evaluate the performance of proposed approach in Section \ref{Performance Evaluation}, and finally, concluding remarks are provided in Section \ref{Conclusion}.
%
%%%%%%%%%%%%%%%%%%%%%%%%%%%%%%%%%%%%%%%%%%%%%%%%%%%%%%%%%%%%%%%%%%%%%%%%
%%%%%%%%%%%%%%%%%%%%%%%%%%%%%%%%%%%%%%%%%%%%%%%%%%%%	BACKGROUND AND RELATED WORK
\section{Background and Related Work}
\label{Background and Related Work}
We start with a brief overview of the relevant aspects of \ac{RL} in Section \ref{Reinforcement Learning}, and then review some of the recent approaches utilizing \ac{RL} on packet routing problem in Section \ref{Related Work: Reinforcement Learning Based Routing}.
%
%%%%%%%%%%%%%%%%%%%%%%%%%%%%%%%%%%%%%%%%%%%%%%%%%%%%	Background: Reinforcement Learning  
\subsection{Background: Reinforcement Learning}
\label{Reinforcement Learning}
Model-free methods enable an agent to learn from  experience while interacting with the environment without prior knowledge of the environment model. The agent at time step $t$ observes the environment state $S_t \in \mathcal S$, takes an action $A_t \in \mathcal A$, gets a reward  $R_{t+1} \in \mathcal R \subset \mathbb R$ at time $t+1$ and the state changes to $S_{t+1}$. In Monte Carlo approaches applied to episodic problems, where the experience is divided into episodes such that all episodes reach a terminal state irrespective of the taken actions, the value estimates are only updated when an episode terminates. While the values can be estimated accurately in these approaches, the learning rate can be dramatically slow. In \ac{TD} learning however, the value estimates are updated incrementally as the agent interacts with the environment without waiting for a final outcome by bootstrapping. Therefore, \ac{TD} learning usually converges faster. In \ac{TD} learning, the goal of the agent is to find the optimal policy $\pi^*$ maximizing the cumulative discounted reward which can be expressed as follows
 \begin{align}
\pi^*=\argmax_{\ \  \ \  \ \pi} \ \mathbb{E}_{\pi} \left[\sum_{k=0}^{\infty} \gamma^k R_{t+k+1} |S_t=s, A_t=a\right],
 \end{align} 
where $0 \leq \gamma \leq 1$ is the discount factor. The Q-learning algorithm \cite{watkins1989learning} is one of the off-policy \ac{TD} control techniques in which the agent learns the optimal action-value function independent of the policy being followed. In Q-learning, the action-value function is updated as follows 
\begin{align}
&Q(S_t, A_t) \leftarrow Q(S_t, A_t)+ \notag \\ &\eta \left( R_{t+1}+ \gamma \max_a Q(S_{t+1},a)-Q(S_t, A_t) \right),
\end{align}
where $\eta$ is the learning rate. We denote by $Y_t$ the \ac{TD} target which is given by
\begin{align}
Y_t = R_{t+1}+ \gamma \max\limits_{a} Q(S_{t+1},a),
\end{align}
and we denote the \ac{TD} error by $\delta_t$ which is expressed as follows
\begin{align}
\delta_t = Y_t-Q(S_t, A_t),
\end{align}
which measures the difference between the old estimate and the \ac{TD} target. Q-learning however suffers from an overestimation problem as the maximization step leads to a significant positive bias that overestimates the actions. Double Q-learning \cite{hasselt2010double} avoids this problem by decoupling the action selection from the action evaluation. In double Q-learning, a $Q$-function finds the maximization action and another $Q$-function estimates the value of taking this action which leads to unbiased estimate of the action-values. Specifically, double Q-learning alternates between updating two $Q$-functions $Q_1$ and $Q_2$ as follows 
\begin{align}
 & Q_1(S_t, A_t) \leftarrow Q_1(S_t, A_t)+ \notag \\ &\eta \left( R_{t+1}+ \gamma  \ Q_2(S_{t+1}, \argmax_a Q_1(S_{t+1}, a) )-Q_1(S_t, A_t) \right), \notag \\
& Q_2(S_t, A_t) \leftarrow Q_2(S_t, A_t)+ \notag \\ &\eta \left( R_{t+1}+ \gamma  Q_1(S_{t+1}, \argmax_a Q_2(S_{t+1}, a) )-Q_2(S_t, A_t) \right).
\end{align}

In large-scale \ac{RL} problems, storing and learning the action-value function for all states and all actions is inefficient. In order to tackle these challenges, the action-value function is usually approximated by representing it as a function parameterized by a weight vector $\bm{\theta}$ instead of using a table. This approximated function can be a linear function in the features of the state, where $\bm{\theta}$ is the feature weights vector, or can be computed for instance using a \ac{DNN}, where $\bm{\theta}$ represents the weights of the neural network. The \ac{DQN} algorithm  \cite{mnih2013playing} extends the tabular Q-learning algorithm by approximating the action-value function and learning a parameterized action-value function instead.
Specifically, an online neural network whose input is the state $s$ outputs the estimated action-value function $Q(s, a; \bm{\theta_t})$ for each action $a \in \mathcal A$ in this state, where $\bm{\theta_t}$ are the parameters of the neural network at time $t$. 
%\begin{figure}[!htb]
%  \centering
%  \includegraphics[scale=0.35]{figures/NeuralNet.eps}
%  \caption{A neural network approximating the state-value function. The input of the neural network is the state and the outputs are the estimated action-value function for each action in this state. \label{NeuralNet} }
%\end{figure}
In \ac{DQN}, a target neural network with outdated parameters $ \bm{\theta_t}^{-}$, that are copied from the online neural network every $\tau$ steps, is used to find the target of the \ac{RL} algorithm. Deep learning algorithms commonly assume the data samples to be independent and correlated data can significantly slow down the learning \cite{halkjaer1997effect}. However, in \ac{RL} the samples are usually correlated. Therefore, an experience replay memory is used in \ac{DQN} to break the correlations. The replay memory stores the agent experience at each time step and a mini-batch of samples that are drawn \emph{uniformly} at random from the replay memory is used to train the neural network. 

In order to overcome the overestimation issue of \ac{DQN}, the \ac{DDQN} algorithm \cite{van2016deep} extends the tabular double Q-learning algorithm by using an online neural network $\bm{\theta_t}$ that finds the greedy action and a target neural network $\bm{\theta_t}^-$ that evaluates this action. Since sampling uniformly from the memory is inefficient, a \emph{\ac{PER}} approach was developed in \cite{schaul2015prioritized} such that the samples are drawn according to their priorities. In \cite{wang2015dueling}, a dueling network architecture was developed that estimates the state-value function and the state-dependent action advantage function separately. In \cite{hessel2018rainbow}, the Rainbow algorithm was developed by combining some of the recent advances in \ac{DRL} including  \ac{DDQN}, \ac{PER}, dueling  \ac{DDQN} among other approaches. 
%
%%%%%%%%%%%%%%%%%%%%%%%%%%%%%%%%%%%%%%%%%%%%%%%%%%%%	Related Work: Reinforcement Learning Based Routing 
\subsection{Related Work: Reinforcement Learning Based Routing}
\label{Related Work: Reinforcement Learning Based Routing}
In a dynamic network, where the availability of the resources and the demand change frequently, a routing algorithm needs to adapt autonomously and use the resources efficiently to satisfy the \ac{QoS} constraints of the different users. Therefore, the routing problem is a natural fit for application of \ac{RL}. \\ There have been extensive research efforts in developing \ac{RL}-based adaptive routing algorithms in the literature \cite{boyan1994packet, kumar1997dual, subramanian1997ants, choi1996predictive}. In \cite{boyan1994packet}, a Q-learning based routing approach known as Q-routing was developed with the objective of minimizing the packet delay in a distributed model in which the nodes act autonomously. In this approach, a node $x$ estimates the time it takes to deliver a packet to a destination node $d$ based on the estimates received from each neighbor $y$ indicating the estimated remaining time for the packet to reach $d$ from $y$. This update is known as the forward exploration. The Q-routing algorithm was shown, through simulations, to significantly outperform the non-adaptive shortest path algorithm. In \cite{kumar1997dual}, a \ac{DRQ} approach was proposed that incorporates an additional update known as the backward exploration update which improves the speed of convergence of the algorithm. In order to address the different \ac{QoS} requirements, the routing algorithm needs to consider other factors such as packet loss and the utilization of the links beside the delay. In \cite{lin2016qos}, an \ac{RL}-based \ac{QoS}-aware routing protocol algorithm was developed for \glspl{SDN}. Based on the transmission delays, queuing delays, packet losses and the utilization of the links, a \ac{RL} routing protocol was developed and shown through simulations to outperform the Q-routing protocol. 

Taking high-dimensional factors into account makes the tabular Q-learning approaches intractable and \Ac{DRL} techniques provide a promising  alternative that can address this issue by approximating the action-value function efficiently. Recently, there has been much interest in using deep learning techniques to address the packet routing  challenges \cite{stampa2017deep, valadarsky2017learning, pham2018deep, mao2018reward, xu2018experience, suarez2019feature, sun2019sinet, mao2019learning}. A \ac{DRL} approach for routing was developed in \cite{stampa2017deep} with the objective of minimizing the delay in the network. In this approach, a single controller finds all the paths of all source-destination pairs given the demand in the network which represents the bandwidth request of each source-destination pair. This approach however results in a complex state and action spaces and does not scale for large networks as it depends on a centralized controller. Moreover, in this approach, the state representation does not capture the \emph{network topology}. Motivated by the high complexity of the state and action representations of the approaches proposed in \cite{valadarsky2017learning, stampa2017deep}, a feature engineering approach has been recently proposed in \cite{suarez2019feature} that only considers some candidate end-to-end paths for each routing request. This proposed representation was shown to outperform the representation of the approaches proposed in \cite{valadarsky2017learning,stampa2017deep} in some use-cases. In attempt to address the difficulties facing the centralized approaches, a distributed multi-agent \ac{DQN}-based \emph{per-packet} routing algorithm was developed in \cite{you2019toward}, where the objective is to minimize the delay in a \emph{fully distributed} fashion. 
%
%%%%%%%%%%%%%%%%%%%%%%%%%%%%%%%%%%%%%%%%%%%%%%%%%%%%%%%%%%%%%%%%%%%%%%%%
%%%%%%%%%%%%%%%%%%%%%%%%%%%%%%%%%%%%%%%%%%%%%%%%%%%%	HIERARCHICAL DRL PATH CALCULATION
\section{Hierarchical Deep Reinforcement Learning Path Calculation}
\label{Problem Formulation}
In this section, we describe our \ac{DRL}-enabled approach to the packet routing problem. We detail our network model and provide a high level overview of our path calculation approach in Section \ref{Network Model}. We formulate the path calculation problem as a \ac{RL} problem in Section \ref{Reinforcement Learning Formulation}. Finally, we present our \ac{DRL}-based path calculation algorithm in Section \ref{Deep Double Q-Routing with Prioritized Experience Replay}.
%
%%%%%%%%%%%%%%%%%%%%%%%%%%%%%%%%%%%%%%%%%%%%%%%%%%%%	Network Model
\subsection{Network Model}
\label{Network Model}
We start with our notations. We use bold fonts for vectors. For a vector $\mathbf x$, we denote the $i$-th element by $\mathbf x[i]$ and the elements $\mathbf x[i], \mathbf x[i+1], \cdots, \mathbf x[j]$ by $\mathbf x[i \cdots j]$. In a graph $\mathcal G=(\mathcal V, \mathcal E)$, where $\mathcal V$ is the set of vertices and $\mathcal E$ is the set of edges, $e[1]$ and $e[2]$ denote the two end points of the edge $e \in \mathcal E$. 

\begin{figure}
  \centering
  \includegraphics[scale=0.32]{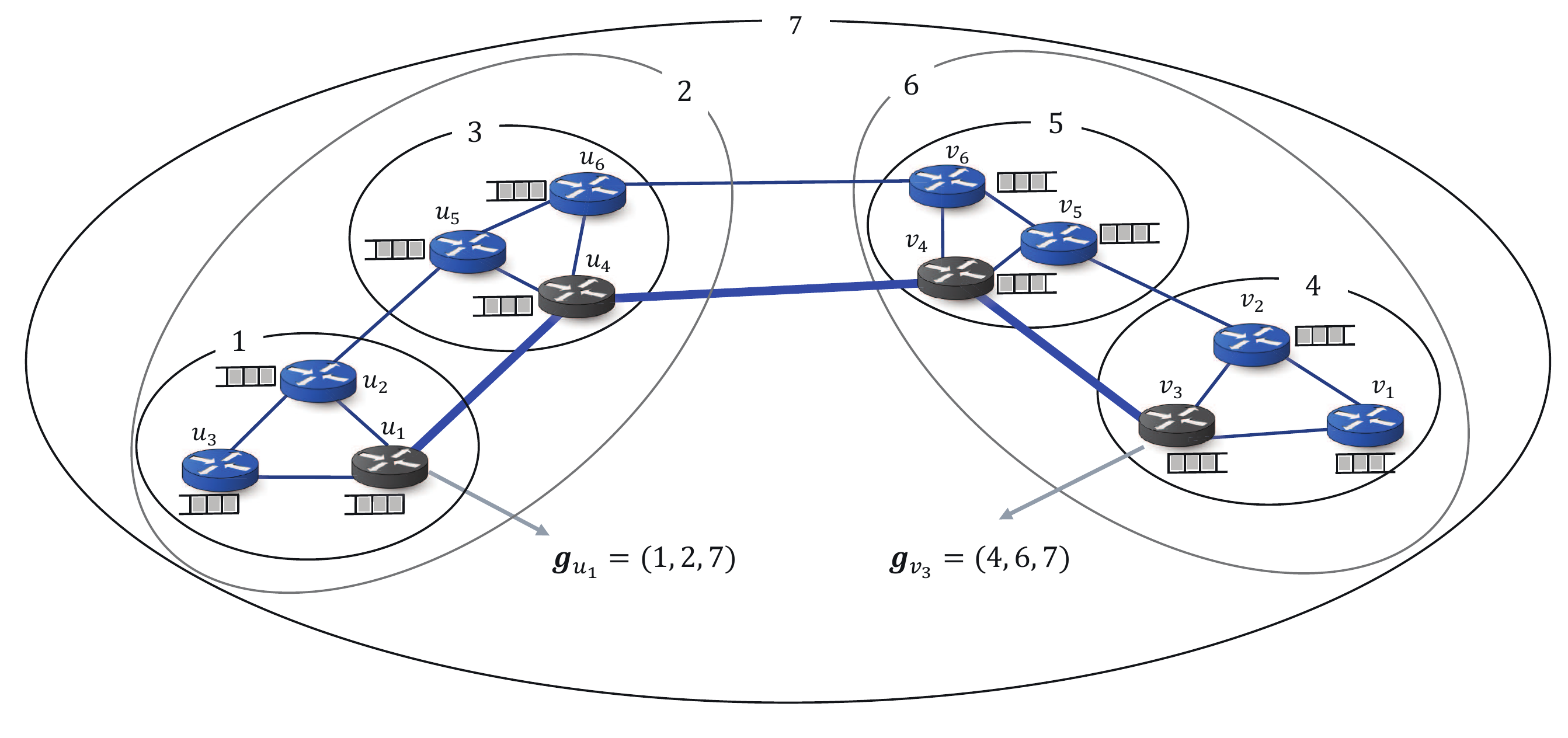}
  \caption{Hierarchical dynamic routing for the case where $H=3$. \label{Clustering} }
\end{figure}

We model the network as a directed graph $\mathcal G=(\mathcal N, \mathcal E)$, where $\mathcal N$ is the set of nodes and $\mathcal E$ is the set of edges representing the links between the nodes. The queuing delay of node $u$ at time $t$ is denoted by $q_u(t)$. A link $e \in \mathcal E$ has a capacity that is denoted by $c_e$. The utilization of a link $e \in \mathcal E$ at time $t$, denoted by $\upsilon_e(t)$, is the ratio between the current rate and the capacity of the link. The transmission delay of a link $e \in \mathcal E$ is denoted by $d_e(t)$ and the packet loss probability is denoted by $p_e(t)$.  The nodes are clustered into groups at $H$ hierarchical levels based on a certain criteria such as latency between the different nodes as shown in Fig. \ref{Clustering}. The $1$-st level is the lowest level in the hierarchy and the $H$-th level is the highest. We refer to a node $v \in \mathcal N$ by its identity and a group vector $\mathbf g_v$ representing how this node is located in the hierarchy. The group vector is of length $H$ and the elements from left to right represent the lowest to the highest level using the identifiers of the groups at those levels. Each group has a designated leader denoted by $l_{\mathbf g}^{(h)}$, where $\mathbf g$ is the group vector of the group leader and $h$ is the group level. 
%%%%%%%%%%%%%%%%%%%%%%%%% Figure:Clustering

A source node $u \in \mathcal N$ that issues a route request to a destination node $v \in \mathcal N$ can find a path $P$ by the assistance of different group leaders at different hierarchical levels. Specifically,  $u$ sends a route request to its local group leader at the $1$-st level. This group leader compares its group vector $\mathbf g_u$ with the group vector of the destination node $\mathbf g_v$ and searches for the highest level in the hierarchy in which they differ denoted by $f_{u, v}$, starting from the highest level. This group leader then sends a route request to the group leader at level $f_{u, v}+1$ requesting a route segment between $\mathbf g_u[f_{u,v}]$ and $\mathbf g_v[f_{u,v}]$. The group leader then finds all possible links $\mathcal E_{\mathbf g_u[f_{u,v}], \mathbf g_v[f_{u,v}]}$ that can connect $\mathbf g_u[f_{u,v}]$ and $\mathbf g_v[f_{u,v}]$ according to the desired \ac{QoS} constraints and chooses a link from these possible links according to the routing policy. This process repeats until a  path $P$ from $u$ to $v$ is calculated. 

In order for the source $u$ to initiate a routing request to $v$ it calls the \texttt{RouteRequest} procedure shown in Algorithm \ref{route request algorithm} with the input tuple $(\mathbf g_{u}, \mathbf g_{v}, H)$ which returns a path $P$ that connects $u$ and $v$. The \texttt{RouteRequest} algorithm depends on two procedures. The first procedure, \texttt{FindLinks}$(\mathbf g_{v_1}[h], \mathbf g_{v_2}[h], h)$, finds all links that can connect $\mathbf g_{v_1}[h]$ and $\mathbf g_{v_2}[h]$ at level $h$ and returns the links among them  corresponding to the desired \ac{QoS} requirements denoted by $\mathcal E_{\mathbf g_{v_1}[h], \mathbf g_{v_2}[h]}$. The second procedure, \texttt{ChooseLink}$(\mathcal E_{\mathbf g_{v_1}[h], \mathbf g_{v_2}[h]})$, returns a link selected from $\mathcal E_{\mathbf g_{v_1}[h], \mathbf g_{v_2}[h]}$. The \texttt{ChooseLink} procedure needs to adapt to the dynamic aspects of the  network such as the utilizations, delays, queue lengths and preferences of the source node. Hence, we design an adaptive \texttt{ChooseLink} procedure based on \ac{DRL} in the next subsection.
%%%%%%%%%%%%%%%%%%%%%%%%% Algorithm:RouteRequest
\begin{algorithm}
  \small
  \caption{RouteRequest ($\mathbf g_{v_1}, \mathbf g_{v_2}, h'$)}\label{route request algorithm}
  \begin{algorithmic}
    \For{$h=h',h'-1, \cdots, 1$}
    \If{$(\mathbf g_{v_1}[h] \neq \mathbf g_{v_2}[h])$}
    \State $l=l_{\mathbf g_{v_1}}^{(h+1)}$ \Comment $l_{\mathbf g_{v_1}}^{(h+1)}=l_{\mathbf g_{v_2}}^{(h+1)}$  
    \State $\mathcal E_{\mathbf g_{v_1}[h], \mathbf g_{v_2}[h]}=\text{FindLinks} (\mathbf g_{v_1}[h], \mathbf g_{v_2}[h], h)$ 
    \State $e=\text{ChooseLink} (\mathcal E_{\mathbf g_{v_1}[h], \mathbf g_{v_2}[h]})$
    \State $P=P \cup {e}$ \Comment{add $e$ to the path $P$}
    \State $u_1=e[1], u_2=e[2]$
   % \If {$(f_{v_1, u_1}<f_{u_1,v_2})$}
    \State $\text{RouteRequest} (\mathbf g_{v_1}, \mathbf g_{u_1}, h-1)$
    \State $\text{RouteRequest} (\mathbf g_{u_2}, \mathbf g_{v_2}, h-1)$
   % \Else
   % \State $\text{RouteRequest} (\mathbf g_{v_1}, \mathbf g_{u_2}, h-1)$ 
   %  \State $\text{RouteRequest} (\mathbf g_{u_1}, \mathbf g_{v_2}, h-1)$
   
%    \EndIf
    \EndIf
    \EndFor
    \Return $P$
  \end{algorithmic}
\end{algorithm}
%
%%%%%%%%%%%%%%%%%%%%%%%%%%%%%%%%%%%%%%%%%%%%%%%%%%%%	Reinforcement Learning Formulation
\subsection{Reinforcement Learning Formulation}
\label{Reinforcement Learning Formulation}
In this subsection, we formulate the link selection problem as a model-free \ac{RL} problem. Each group leader is a \ac{DRL} agent that observes \emph{local information} and \emph{feedback of the source nodes} and acts accordingly. We now describe the state space $\mathcal S$, the action space $\mathcal A$ and our composite reward design approach.  

\textbf{\emph{State space.}} The state of a group leader $l$ at time $t$, $S_t \in \mathcal S$, consists of the partial group vectors $(\mathbf g_{v_1} [k \cdots h], \mathbf g_{v_2} [k \cdots h])$ which represent the network topology, where $k \geq 1$ is a parameter that can be chosen based on memory constraints of the routing nodes. The state also includes utilization, queuing delay and transmission delays of the set of possible links $\mathcal E_{\mathbf g_{v_1}[h], \mathbf g_{v_2}[h]}$. That is, the state is given by
\begin{align}
s_t= (&\mathbf g_{v_1} [k \cdots h], \mathbf g_{v_2} [k \cdots h], q_l(t), \notag \\ &\{\upsilon_e(t), d_e(t)\}_{e \in \mathcal E_{\mathbf g_{v_1}[h], \mathbf g_{v_2}[h]}}).
\end{align}

\textbf{\emph{Action space.}} The action $A_t \in \mathcal A$ of the group leader $l$ represents the link $e \in \mathcal E_{ \mathbf g_{v_1}[h], \mathbf g_{v_2}[h]}$ that the group leader chooses for the routing request.

\textbf{\emph{Reward.}} The group leader $l$ at state $S_t$ that takes an action $A_t \in \mathcal A$ gets a reward $R_{t+1} \in \mathcal R$. We design a composite reward \cite{mao2018reward} such that it captures both the global and the local aspects of the path calculation problem as follows. 
\begin{itemize}
  \item \textbf{Global Reward.}  The group leader takes actions based purely on the local information available, without knowing the effect of these actions on the end-to-end \ac{QoS}-type constraints. Hence, we design a global reward control signal to address this issue that is assigned to all group leaders involved in the routing of a certain flow without distinction. Specifically, a source node $u$ that issues a route request to a destination node $v$ sends a control signal $y_{u, v}$ to the group leaders involved in the routing that is between $0$ and $1$ indicating the satisfaction of the source about the selected path. We note that the global reward signal may not be instantaneous as the source does not continuously send it. Instead, the source sends it from time to time and we assume the group leaders receive it $T$ steps after choosing the path. That is, the global reward of the group leader that selects a link $e \in \mathcal E_{\mathbf g_{v_1}[h], \mathbf g_{v_2}[h]}$ in state $s_t$ is expressed as follows
  \begin{align}
  r_{t+T}^G=w_1 y_{u, v},
  \end{align}
  where $w_1 \geq 0$ is the weight of the global reward. 
  
  \item \textbf{Local Reward.} The local reward is assigned individually to each group leader based on the individual contribution. Specifically, the local reward depends on the queuing delay, transmission delay, packet loss  of the selected link and how well is the group leader balancing the load over possible links. The local reward is expressed as follows
  \begin{align}
 & r_{t+1}^{L}=w_2 \ \mathbf 1(\upsilon_e (t) \leq  \upsilon_{\rm {th}}) - w_3 \left(q_l(t)+d_e(t) \right)- \notag \\ & w_4 \ p_e(t) +w_5 \exp{ ( -\textstyle \sum_{e' \in \mathcal E_{\mathbf g_{v_1}[k], \mathbf g_{v_2}[k]}} |\upsilon_{e'}(t)-\overline \upsilon(t)| )},
  \end{align}
  where $w_2, w_3, w_4, w_5 \geq 0$ are weights that can be chosen by the network operator, $\upsilon_{\rm {th}}$ denotes a utilization threshold that it is undesirable to exceed and $\overline \upsilon(t)$ denotes the average utilization of links in $\mathcal E_{\mathbf g_{v_1}[h], \mathbf g_{v_2}[h]}$ which is expressed as follows
  \begin{align}
  \overline \upsilon(t)=\frac{1}{|\mathcal E_{\mathbf g_{v_1}[h], \mathbf g_{v_2}[h]}|} \textstyle \sum\limits
  _{e' \in \mathcal E_{\mathbf g_{v_1}[h], \mathbf g_{v_2}[h]}} \upsilon_{e'}(t).
  \end{align}
\end{itemize}
Therefore, the composite reward combining global and local rewards is given by
\begin{align}
r_{t}=r_{t}^G + r_{t}^L.
\end{align}
%
%%%%%%%%%%%%%%%%%%%%%%%%%%%%%%%%%%%%%%%%%%%%%%%%%%%%	Deep Double Q-Routing with Prioritized Experience Replay
\subsection{Deep Double Q-Routing with Prioritized Experience Replay}
\label{Deep Double Q-Routing with Prioritized Experience Replay}
%%%%%%%%%%%%%%%%%%%%%%%%% Figure:DeepRLSchematic
\begin{figure*}[!htb]
  \centering
  \hspace*{3cm}\includegraphics[scale=0.65]{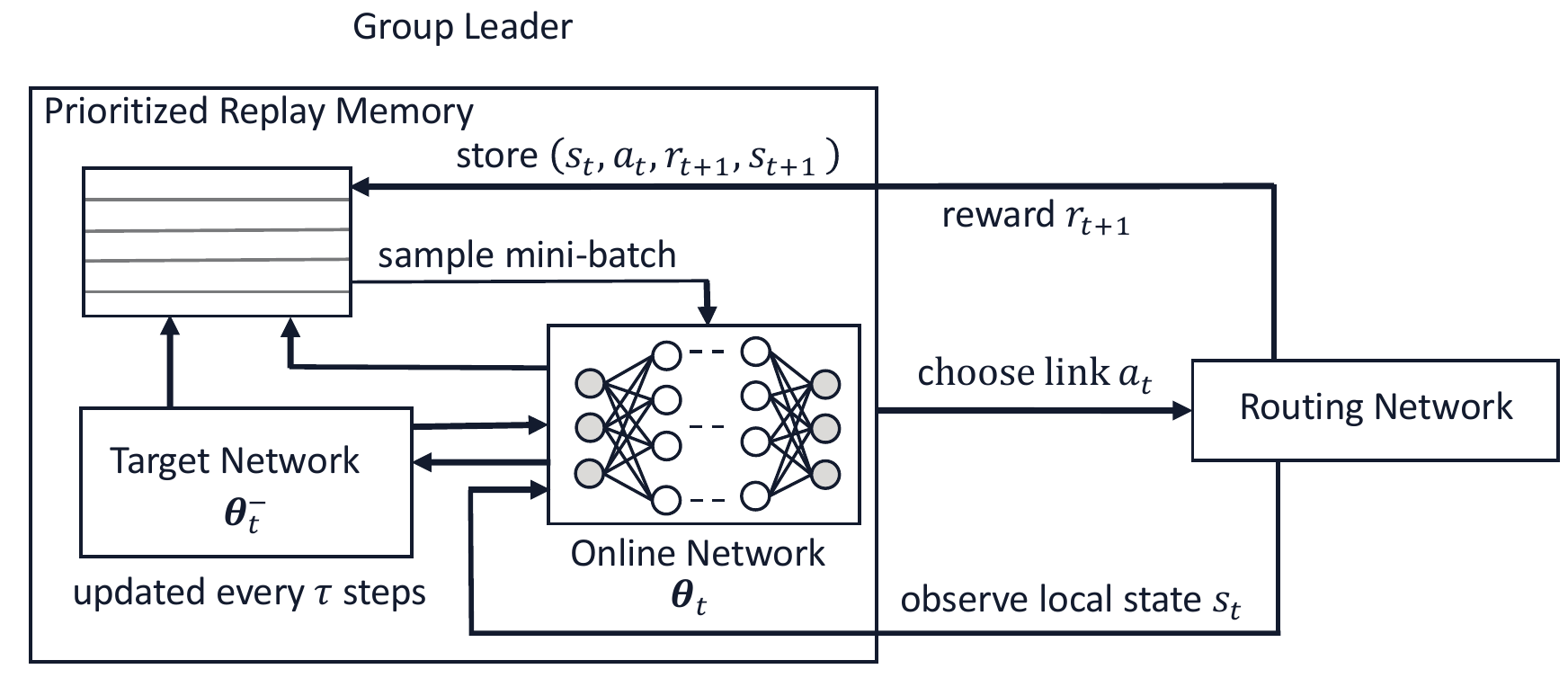}
  \caption{The learning architecture of the group leader. \label{DeepRLSchematic}}
\end{figure*}
We consider a model-free, off-policy algorithm based on \ac{DDQN} algorithm developed in \cite{van2016deep} as depicted in Fig. \ref{DeepRLSchematic} and given in Algorithm \ref{Q-learning choose link algorithm}. As we have explained in Section \ref{Reinforcement Learning}, the action selection uses an online neural network with weights $\bm{\theta}_t$, referred to as Q-network, to estimate the action-value function. The input of neural network is the state of group leader $S_t$ and the outputs are the estimated action-value function $Q(S_t, a; \bm{\theta}_t)$ for each action $a \in \mathcal A$ in that given state, where each output unit corresponds to a particular action (link). The action evaluation uses a target network with weights $\bm{\theta_t} ^ {-}$, which is a copy of the online network weights that is updated every $\tau$ steps as $\bm{\theta_t}^{-}=\bm{\theta}_t$. The target is expressed as follows
\begin{align}
Y_t=R_{t+1}+ \gamma Q(S_{t+1}, \argmax_a Q(S_{t+1}, a; \bm{\theta_t}); \bm{\theta_t}^{-}), 
\end{align}
and the parameters $\bm{\theta}_t$ are updated as follows
\begin{align}
\bm{\theta}_{t+1}= \bm{\theta}_t+ \eta (Y_t - Q(S_t, A_t; \bm{\theta}_t)) \nabla_{\bm{\theta}_t} Q(S_t, A_t; \bm{\theta}_t).
\end{align}

We use experience replay memory $\mathcal M$ of size $M$ to store the  experience of a group leader at each time step $t$ as $(s_t, a_t, r_{t+1}, s_{t+1})$. These experiences are replayed later at a rate that is based on their priority, where the priority depends on how surprising was that experience as indicated by its \ac{TD}-error. Specifically, we use a proportional priority memory where the priority of transition $i$ is expressed as follows 
\begin{align}
p_i=|\delta_i|+ \epsilon',
\end{align}
where $\delta_i$ is the \ac{TD}-error of transition $i$ and $\epsilon'>0$ is a constant. The probability of sampling a transition $i$ from $\mathcal M$ is given by
\begin{align}
P_i=p_i^\alpha / \sum\limits_j p_j^\alpha,
\end{align}
where $\alpha$ is a parameter that controls the level of prioritization and $\alpha=0$ corresponds to the uniform sampling case. Sampling experiences based on the priority introduces a bias, and \ac{IS} can correct this bias. In particular, the \ac{IS} weight of transition $i$ is given by
\begin{align}
w_i= \left( \frac{1}{M} \ . \ \frac{1}{P_i} \right)^\beta,
\end{align}
where $\beta$ is defined by a schedule that starts with an initial value $\beta_0$ and reaches $1$ at the end of learning. These weights are normalized by the maximum weight $\max_i w_i$.
%%%%%%%%%%%%%%%%%%%%%%%%% Algorithm:ChooseLink
\begin{algorithm}
  \small
  \caption{ChooseLink $(\mathcal E_{\mathbf g_{v_1}[h], \mathbf g_{v_2}[h]})$ \label{Q-learning choose link algorithm}}
  \begin{algorithmic}[!htb]
    \State \textbf{Parameters}: Replay memory size $M$, mini-batch size $m$, replay period $K$, exponents $\alpha, \beta$, discount factor $\gamma$, learning rate $\eta$, exploration rate $\epsilon$, update period $\tau$
    \State $\mathcal A \leftarrow \mathcal E_{\mathbf g_{v_1}[h], \mathbf g_{v_2}[h]}$ 
     \State Observe current state $s_t$ and reward $r_t$
    \State Store transition $(s_{t-1}, a_{t-1}, r_t, s_{t})$ in $\mathcal M$ with priority $p_t= \max_{i<t} p_i$
    \If{$t \mod \tau== 0$}
    \State Update the target network weights  $\bm{\theta}_t^{-} \leftarrow \bm{\theta}_t$
    \EndIf
    \If {$t \mod  K ==0$} 
    \State Initialize $\Delta=0$ \Comment $\Delta$ is used to update $\bm{\theta}_t$
    \For{$j=1, 2, \cdots, m$}
    \State Sample transition $j$
    \begin{align*}
     (s_j, a_j, r_{j+1}, s_{j+1}) \sim P_j= p_j^ \alpha / \sum_i p_i^\alpha
      \end{align*}
    \State Compute target 
    \begin{align*}
    & ~~~~~~ ~~~~~~~~~y_{j+1}=r_{j+1} + \gamma Q(s_{j+1}, \argmax_a Q(s_{j+1}, a; \bm{\theta_t}); \bm{\theta_t}^{-})
    \end{align*}
    \State Compute \ac{IS} weight $w_j=(M P_j)^{-\beta}$
    \State Normalize \ac{IS} weight 
    $w_j \leftarrow w_j /\max_i w_i$
    \State Compute TD-error $
    \delta_j= y_{j+1}-Q(s_j, a_j; \bm{\theta}_t)$
    \State Update priority $p_j \leftarrow |\delta_j|$
    \State Update the weight-change 
    \begin{align*}
    \Delta \leftarrow \Delta+ w_j. \delta_j 
     \nabla_{\bm{\theta}_t} Q(s_j, a_j; \bm{\theta}_t)
     \end{align*}
    \EndFor
    \State Update online weights $\bm{\theta}_{t+1} \leftarrow \bm{\theta}_t + \eta. \Delta$
    \EndIf
      \State Choose an action (link) $a_t$ from $s_t$  $\epsilon$-greedily as follows
    \begin{align}
    \pi (a_t|s_t)=
    \begin{cases}
    1-\epsilon+\epsilon/|\mathcal A|  \ \ \text{if $a_t=\argmax\limits_a  Q(s_t,a; \bm{\theta}_t)$}, \\
    \epsilon/|\mathcal A| \ \  \text{otherwise.}
    \end{cases}
    \end{align}
  \State $t \leftarrow t+1$ \Comment{$t=0$ initially} \\
  \Return the selected link 
  \end{algorithmic}
\end{algorithm}
%
%%%%%%%%%%%%%%%%%%%%%%%%%%%%%%%%%%%%%%%%%%%%%%%%%%%%%%%%%%%%%%%%%%%%%%%%
%%%%%%%%%%%%%%%%%%%%%%%%%%%%%%%%%%%%%%%%%%%%%%%%%%%%	PERFORMANCE EVALUATION
\section{Performance Evaluation}
\label{Performance Evaluation}
\begin{figure*}[!htb]
	\centering
	\includegraphics[width=0.7\textwidth,height=0.33\textwidth]{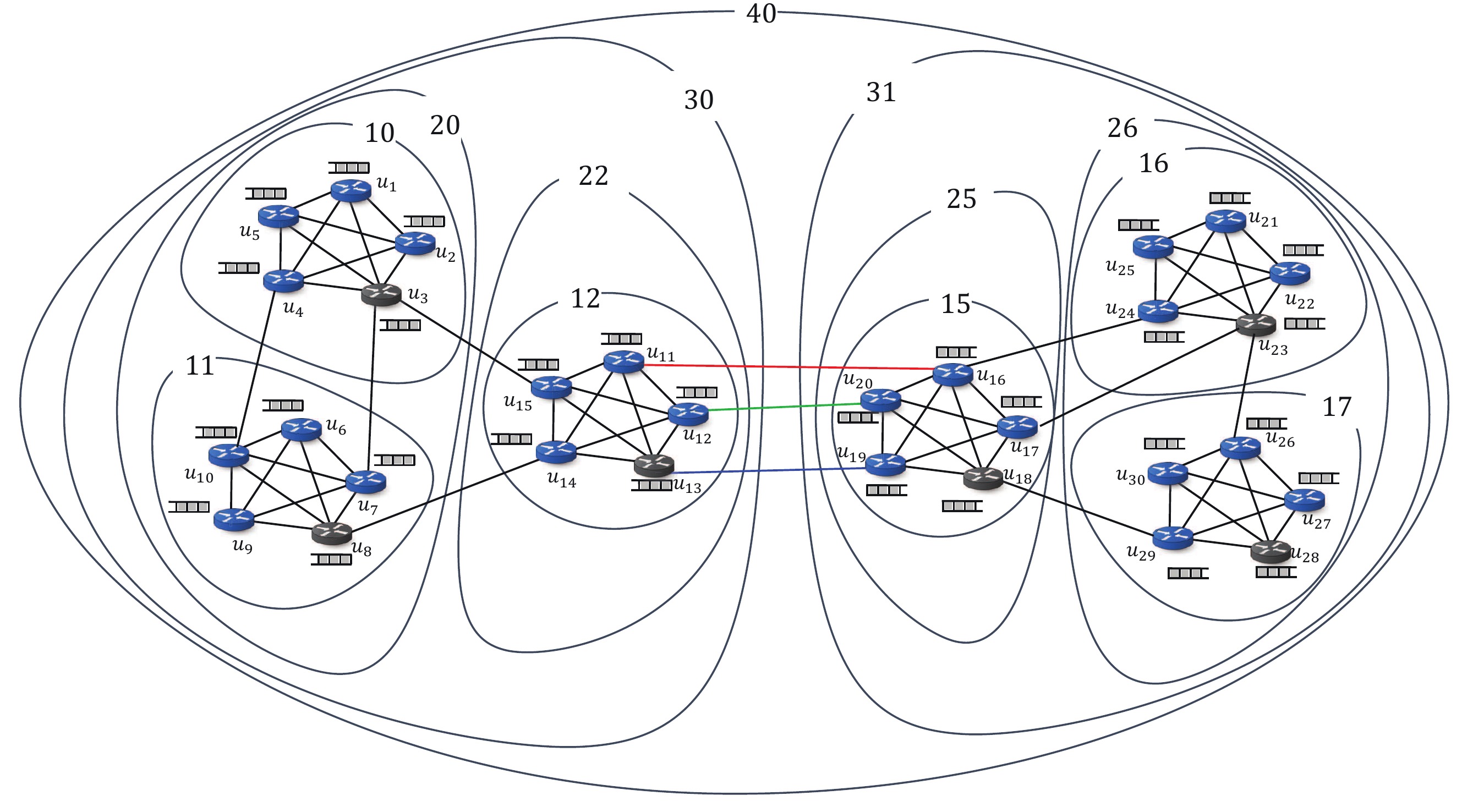}
	\caption{The topology considered in our experiment setup.\label{SetUp}}
\end{figure*}

\begin{figure}[!htb]
	\centering
	\includegraphics[scale=0.475]{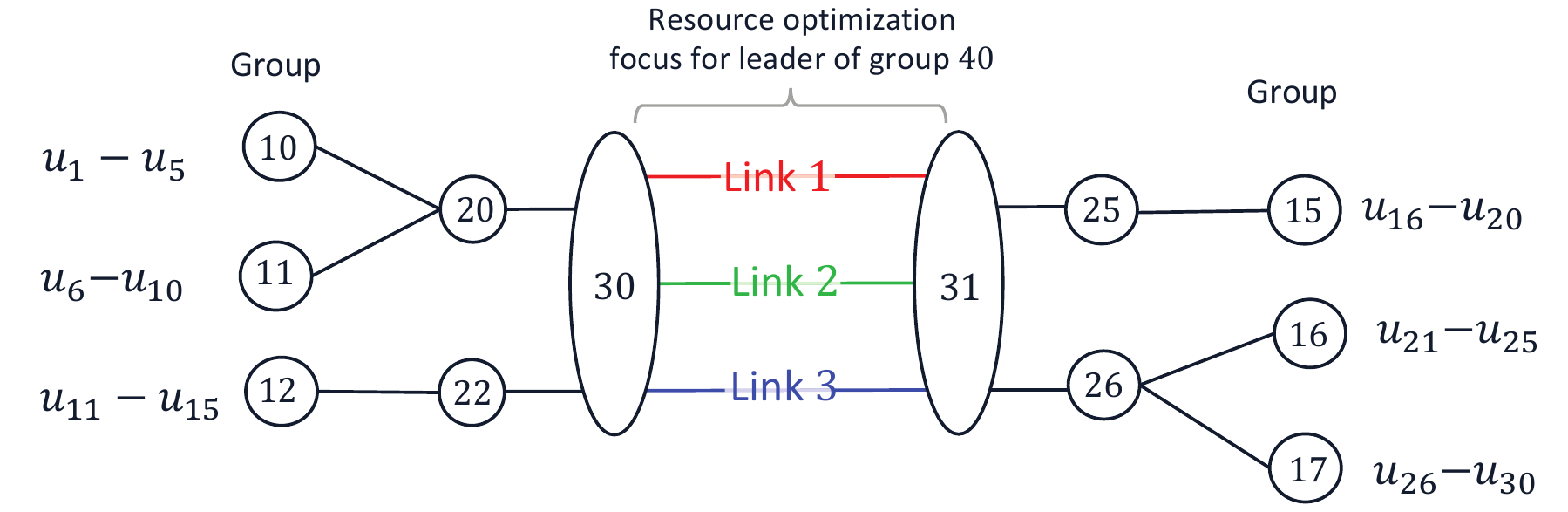}
	\caption{A simplified schematic for the topology of our experiment setup.\label{SetUP_Simplified}}
\end{figure}

In this section, we evaluate the performance of our approach on OpenAI Gym \cite{brockman2016openai} considering the topology shown in Fig. \ref{SetUp} and Fig. \ref{SetUP_Simplified}. We consider a dense neural network for the agent with $2$ hidden layers, RMSprop optimizer \cite{hinton2012neural} and Huber loss. The hidden layers have $32$ neurons each and have \ac{ReLU} activation function. The output layer has a linear activation function. We have selected our parameters as shown in Table \ref{table:hyperparameters} and Table \ref{table:network_parameters}. 
%%%%%%%%%%%%%%%%%%%%%%%%% Figure:SetUp
%%%%%%%%%%%%%%%%%%%%%%%%% Figure:SetUP_Simplified
%%%%%%%%%%%%%%%%%%%%%%%%% Table:hyperparameters

\begin{table}[!htb]
	\caption{The hyperparameters used in our experiment.}
	\centering % used for centering table
	\begin{tabular}{c c}     
		{\bf Parameter} & {\bf Value} \\ [0.5ex] % inserts table
		%heading
		\hline \hline % inserts single horizontal line
		Replay memory size  $M$ & $1000$   \\ % inserting body of the table
		Mini-batch size $m$ & $32$ \\
		Replay period $K$ & $1$  \\
		Update period $\tau$ & $250$ \\
		Priortization exponent $\alpha$ & 0.5 \\
		Importance sampling exponent $\beta$ & linearly annealed from  $0.4$ to $1$\\
		Minimum exploration rate $\epsilon_{\mathrm{min}}$ & $0.01$ \\
		Maximum exploration rate $\epsilon_{\mathrm{max}}$ & $1$ \\
		Exploration rate exponent $\lambda$ & $0.01$ \\
		Exploration rate $\epsilon$ & $\epsilon_{\mathrm{min}}+(\epsilon_{\mathrm{max}}-\epsilon_{\mathrm{min}}) \ e^{- \lambda t}$ \\
	    Prioritization constant $\epsilon'$	& $0.01$ \\
		[1ex] 
		\hline %inserts single line
	\end{tabular}
	\label{table:hyperparameters} % is used to refer this table in the text
\end{table}
%%%%%%%%%%%%%%%%%%%%%%%%% Table:network_parameters
\begin{table}[!htb]
	\caption{Network simulation parameters. } % title of Table
	\centering % used for centering table
	\begin{tabular}{c c}      
		{\bf Parameter} & {\bf Value} \\ [0.5ex] % inserts table
		%heading
		\hline \hline % inserts single horizontal line
		Global reward weights  $w_1$ & $2$   \\ % inserting body of the table
		Utilization threshold weight  $w_2$ & $10$ \\
		Delay weight $w_3$ & $0$ \\
		Packet loss weight $w_4$ & $0$ \\
		Load balancing weight $w_5$ & $20$  \\
		Link capacity  & $70$ flow per second  \\
		Utilization threshold $\upsilon_{\rm {th}}$ & $0.79$ \\
		Minimum flow duration & $6$ s \\
		Maximum flow duration & $10$ s \\
		Requests inter-arrival time & $3 \ \mu s$  \\
		[1ex] 
		\hline %inserts single line
	\end{tabular}
	\label{table:network_parameters} % is used to refer this table in the text
\end{table}

As a matter of showcase, we pick the source-destination pair $(u_1, u_{16})$ and focus mainly on the utilization components of our reward. The source generates random flow requests with random durations and  the group leader of group $40$ selects a link from link $1$, link $2$ and link $3$ to connect group $30$ and group $31$. Therefore, based on which link is selected, we consider $3$ paths between this source-destination pair. We assume that the source prefers the path that involves link $1$, then prefers the path that involves link $2$, then the path through link $3$. Specifically, the source assigns a global reward of $1$ to the first path, global reward of $0.5$ to the second path and a global reward of $0.3$ to the third path.

\begin{figure}[!htb]
	\centering
	\includegraphics[scale=0.45]{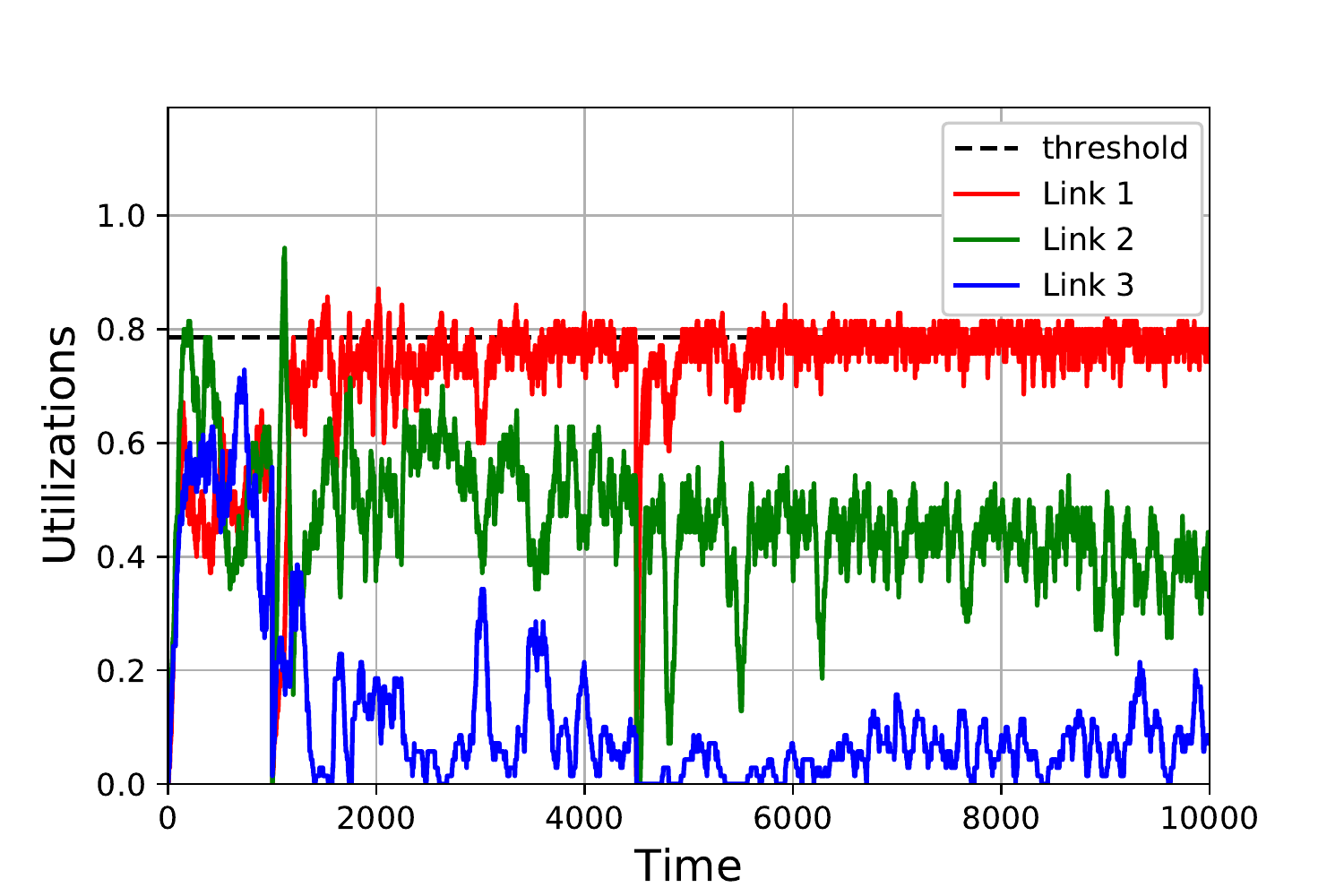}
	\caption{The utilization of the three links of group leader $40$. \label{fig:utilizations} }
\end{figure}

\begin{figure}[!htb]
	\centering
	\includegraphics[scale=0.45]{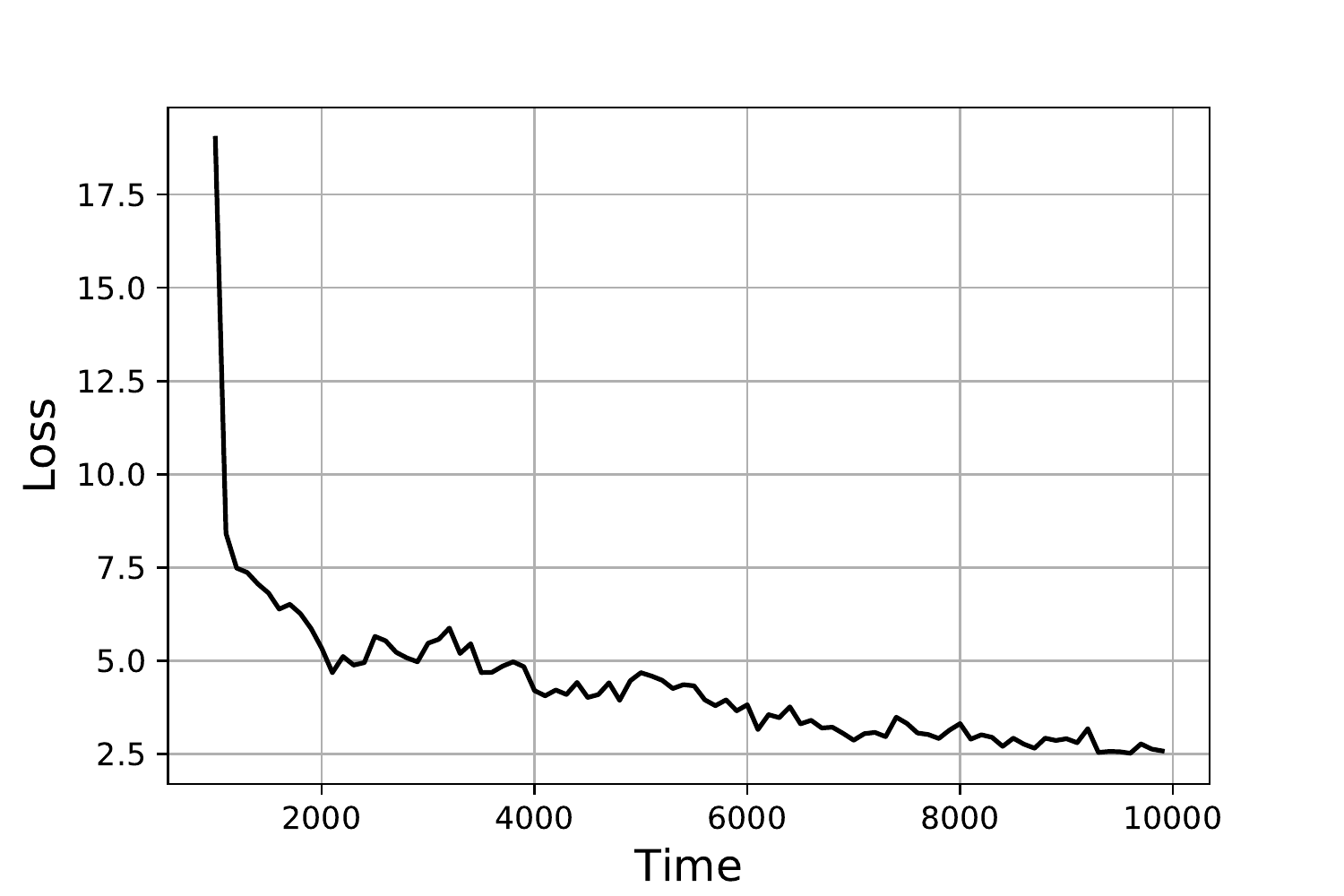}
	\caption{Loss as a function of time. \label{fig:loss} }
\end{figure}

\begin{figure}[!htb]
	\centering
	\includegraphics[scale=0.45]{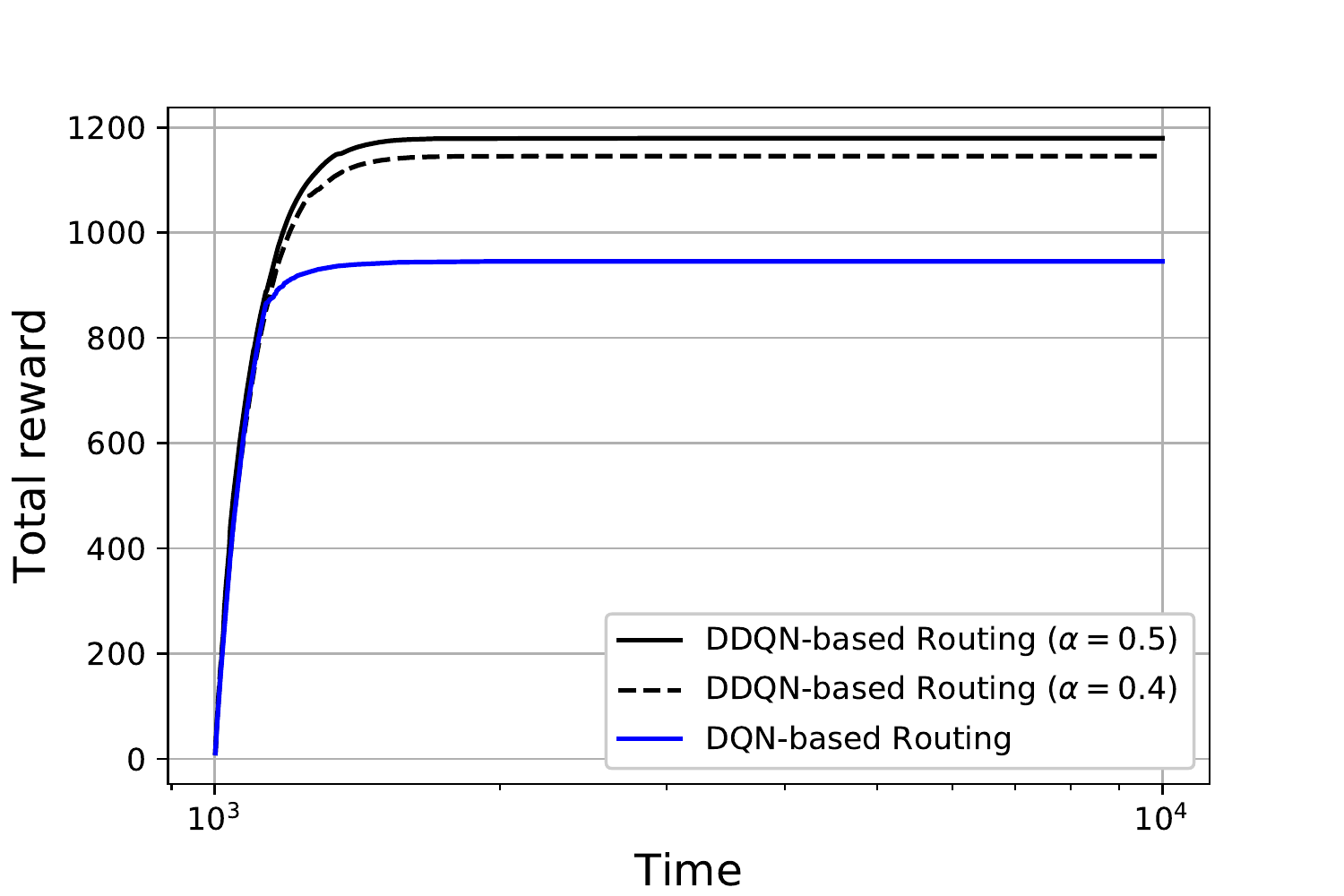}
	\caption{Total discounted reward of the DDQN and the DQN-based routing. \label{fig:comparison} }
\end{figure}

In Fig. \ref{fig:utilizations}, we show the utilizations of three links that group leader $40$ manages. In the first $M$ steps, the group leader initializes the replay memory by selecting links at random and after that it starts to learn. We observe that the group leader learns the preferences of the source node: the first path, the second, then the third. Moreover, the group leader selects links in a way that the utilization of links do not exceed the predefined utilization threshold $\upsilon_{\rm {th}}$ and importantly the group leader balances the load across these three links while considering the preferences of the source. Specifically, the group leader picks the first link more than the other two links until the utilization of this link reaches the predefined threshold. The group leader then selects the second link more than the third link, but we observe that the group leader still selects the third link for some of the requests, even before utilization of the second link reaches the predefined threshold, to balance the load across these three links. In Fig. \ref{fig:loss}, we also show the neural network loss of the group leader during online learning process, highlighting a better performance over time. Finally, we compare the total discounted reward of the \ac{DQN}-based routing algorithm with the \ac{DDQN}-based routing under the same traffic pattern  in Fig. \ref{fig:comparison}. As expected, we notice that the \ac{DDQN}-based routing results in a higher total reward as compared with the \ac{DQN}-based routing.
%%%%%%%%%%%%%%%%%%%%%%%%% Figure:utilizations
%%%%%%%%%%%%%%%%%%%%%%%%% Figure:loss
%%%%%%%%%%%%%%%%%%%%%%%%% Figure:comparison
%
%%%%%%%%%%%%%%%%%%%%%%%%%%%%%%%%%%%%%%%%%%%%%%%%%%%%%%%%%%%%%%%%%%%%%%%%
%%%%%%%%%%%%%%%%%%%%%%%%%%%%%%%%%%%%%%%%%%%%%%%%%%%%	CONCLUSION
\section{Conclusion}
\label{Conclusion}
In this paper, we presented a hierarchical approach to the packet routing problem based on the \ac{DDQN} algorithm. Our approach is expected to scale in large networks as the path calculation is hierarchically distributed over designated nodes in the network rather than a centralized node calculating the paths for all nodes in the network. Moreover, our path calculation algorithm can adapt dynamically to rapid changes in the network and utilizes the resources based on a policy to determine real-time paths. Our future work includes assessment of our algorithm in large-scale network topologies and quantifying the advantages in terms of the routing performance, memory requirements and communication efficiency.
%
%%%%%%%%%%%%%%%%%%%%%%%%%%%%%%%%%%%%%%%%%%%%%%%%%%%%%%%%%%%%%%%%%%%%%%%%
%%%%%%%%%%%%%%%%%%%%%%%%%%%%%%%%%%%%%%%%%%%%%%%%%%%%	REFERENCES
\bibliographystyle{IEEEtran}
\bibliography{IEEEabrv,Routing}

\begin{thebibliography}{10}
\providecommand{\url}[1]{#1}
\csname url@samestyle\endcsname
\providecommand{\newblock}{\relax}
\providecommand{\bibinfo}[2]{#2}
\providecommand{\BIBentrySTDinterwordspacing}{\spaceskip=0pt\relax}
\providecommand{\BIBentryALTinterwordstretchfactor}{4}
\providecommand{\BIBentryALTinterwordspacing}{\spaceskip=\fontdimen2\font plus
\BIBentryALTinterwordstretchfactor\fontdimen3\font minus
  \fontdimen4\font\relax}
\providecommand{\BIBforeignlanguage}[2]{{%
\expandafter\ifx\csname l@#1\endcsname\relax
\typeout{** WARNING: IEEEtran.bst: No hyphenation pattern has been}%
\typeout{** loaded for the language `#1'. Using the pattern for}%
\typeout{** the default language instead.}%
\else
\language=\csname l@#1\endcsname
\fi
#2}}
\providecommand{\BIBdecl}{\relax}
\BIBdecl

\bibitem{di1998efficient}
M.~Di~Ianni, ``Efficient delay routing,'' \emph{Theoretical Computer Science},
  vol. 196, no. 1-2, pp. 131--151, 1998.

\bibitem{filsfils2015segment}
C.~Filsfils, N.~K. Nainar, C.~Pignataro, J.~C. Cardona, and P.~Francois, ``The
  segment routing architecture,'' in \emph{2015 IEEE Global Communications
  Conference (GLOBECOM)}, pp. 1--6.

\bibitem{ash2006path}
J.~Ash and A.~Farrel, ``A path computation element ({PCE})-based
  architecture,'' \emph{IETF, RFC4655}, August 2006.

\bibitem{watkins1989learning}
C.~J. C.~H. Watkins, ``Learning from delayed rewards,'' \emph{PhD thesis,
  King’s College, Oxford}, 1989.

\bibitem{hasselt2010double}
H.~V. Hasselt, ``Double q-learning,'' in \emph{Advances in Neural Information
  Processing Systems}, 2010, pp. 2613--2621.

\bibitem{mnih2013playing}
V.~Mnih, K.~Kavukcuoglu, D.~Silver, A.~Graves, I.~Antonoglou, D.~Wierstra, and
  M.~Riedmiller, ``Playing atari with deep reinforcement learning,''
  \emph{arXiv preprint arXiv:1312.5602}, 2013.

\bibitem{halkjaer1997effect}
S.~Halkj{\ae}r and O.~Winther, ``The effect of correlated input data on the
  dynamics of learning,'' in \emph{Advances in neural information processing
  systems}, 1997, pp. 169--175.

\bibitem{van2016deep}
H.~Van~Hasselt, A.~Guez, and D.~Silver, ``Deep reinforcement learning with
  double q-learning,'' in \emph{Thirtieth AAAI Conference on Artificial
  Intelligence}, 2016.

\bibitem{schaul2015prioritized}
T.~Schaul, J.~Quan, I.~Antonoglou, and D.~Silver, ``Prioritized experience
  replay,'' \emph{arXiv preprint arXiv:1511.05952}, 2015.

\bibitem{wang2015dueling}
Z.~Wang, T.~Schaul, M.~Hessel, H.~Hasselt, M.~Lanctot, and N.~Freitas,
  ``Dueling network architectures for deep reinforcement learning,'' in
  \emph{International Conference on Machine Learning}, 2016, pp. 1995--2003.

\bibitem{hessel2018rainbow}
M.~Hessel, J.~Modayil, H.~Van~Hasselt, T.~Schaul, G.~Ostrovski, W.~Dabney,
  D.~Horgan, B.~Piot, M.~Azar, and D.~Silver, ``Rainbow: Combining improvements
  in deep reinforcement learning,'' in \emph{Thirty-Second AAAI Conference on
  Artificial Intelligence}, 2018.

\bibitem{boyan1994packet}
J.~A. Boyan and M.~L. Littman, ``Packet routing in dynamically changing
  networks: A reinforcement learning approach,'' in \emph{Advances in neural
  information processing systems}, 1994, pp. 671--678.

\bibitem{kumar1997dual}
S.~Kumar and R.~Miikkulainen, ``Dual reinforcement q-routing: An on-line
  adaptive routing algorithm,'' in \emph{Proceedings of the artificial neural
  networks in engineering Conference}, 1997, pp. 231--238.

\bibitem{subramanian1997ants}
D.~Subramanian, P.~Druschel, and J.~Chen, ``Ants and reinforcement learning: A
  case study in routing in dynamic networks,'' in \emph{IJCAI (2)}.\hskip 1em
  plus 0.5em minus 0.4em\relax Citeseer, 1997, pp. 832--839.

\bibitem{choi1996predictive}
S.~P. Choi and D.-Y. Yeung, ``Predictive q-routing: A memory-based
  reinforcement learning approach to adaptive traffic control,'' in
  \emph{Advances in Neural Information Processing Systems}, 1996, pp. 945--951.

\bibitem{lin2016qos}
S.-C. Lin, I.~F. Akyildiz, P.~Wang, and M.~Luo, ``Qos-aware adaptive routing in
  multi-layer hierarchical software defined networks: A reinforcement learning
  approach,'' in \emph{2016 IEEE International Conference on Services Computing
  (SCC)}.\hskip 1em plus 0.5em minus 0.4em\relax IEEE, 2016, pp. 25--33.

\bibitem{stampa2017deep}
G.~Stampa, M.~Arias, D.~Sanchez-Charles, V.~Munt{\'e}s-Mulero, and A.~Cabellos,
  ``A deep-reinforcement learning approach for software-defined networking
  routing optimization,'' \emph{arXiv preprint arXiv:1709.07080}, 2017.

\bibitem{valadarsky2017learning}
A.~Valadarsky, M.~Schapira, D.~Shahaf, and A.~Tamar, ``Learning to route with
  deep rl,'' in \emph{NIPS Deep Reinforcement Learning Symposium}, 2017.

\bibitem{pham2018deep}
T.~A.~Q. Pham, Y.~Hadjadj-Aoul, and A.~Outtagarts, ``Deep reinforcement
  learning based qos-aware routing in knowledge-defined networking,'' in
  \emph{International Conference on Heterogeneous Networking for Quality,
  Reliability, Security and Robustness}.\hskip 1em plus 0.5em minus 0.4em\relax
  Springer, 2018, pp. 14--26.

\bibitem{mao2018reward}
H.~Mao, Z.~Gong, and Z.~Xiao, ``Reward design in cooperative multi-agent
  reinforcement learning for packet routing,'' 2018.

\bibitem{xu2018experience}
Z.~Xu, J.~Tang, J.~Meng, W.~Zhang, Y.~Wang, C.~H. Liu, and D.~Yang,
  ``Experience-driven networking: A deep reinforcement learning based
  approach,'' in \emph{IEEE INFOCOM 2018-IEEE Conference on Computer
  Communications}, pp. 1871--1879.

\bibitem{suarez2019feature}
J.~Suarez-Varela, A.~Mestres, J.~Yu, L.~Kuang, H.~Feng, P.~Barlet-Ros, and
  A.~Cabellos-Aparicio, ``Feature engineering for deep reinforcement learning
  based routing,'' in \emph{2019 IEEE International Conference on
  Communications (ICC)}, pp. 1--6.

\bibitem{sun2019sinet}
P.~Sun, J.~Li, Z.~Guo, Y.~Xu, J.~Lan, and Y.~Hu, ``Sinet: Enabling scalable
  network routing with deep reinforcement learning on partial nodes,'' in
  \emph{Proceedings of the ACM SIGCOMM 2019 Conference}, 2019, pp. 88--89.

\bibitem{mao2019learning}
H.~Mao, Z.~Gong, Z.~Zhang, Z.~Xiao, and Y.~Ni, ``Learning multi-agent
  communication under limited-bandwidth restriction for internet packet
  routing,'' \emph{arXiv preprint arXiv:1903.05561}, 2019.

\bibitem{you2019toward}
X.~You, X.~Li, Y.~Xu, H.~Feng, and J.~Zhao, ``Toward packet routing with
  fully-distributed multi-agent deep reinforcement learning,'' \emph{IEEE
  RAWNET workshop, WiOpt 2019, Avignon, France}.

\bibitem{brockman2016openai}
G.~Brockman, V.~Cheung, L.~Pettersson, J.~Schneider, J.~Schulman, J.~Tang, and
  W.~Zaremba, ``Open{AI} {G}ym,'' \emph{arXiv preprint arXiv:1606.01540}, 2016.

\bibitem{hinton2012neural}
G.~Hinton, N.~Srivastava, and K.~Swersky, ``{RMS}prop: Divide the gradient by a
  running average of its recent magnitude,'' \emph{Lecture notes on Neural
  networks for machine learning}.

\end{thebibliography}
\end{document}